\documentclass{aa}
\usepackage{natbib}
\usepackage{graphicx}
\usepackage{amssymb}
\usepackage{color}

\begin{document}

\title{An alternative model for the origin of gaps in circumstellar disks}

\author{Eduard I. Vorobyov \inst{1,2}, Zsolt Regaly \inst{3},  Manuel Guedel \inst{1} and D. N. C. Lin \inst{4}}
\authorrunning{Vorobyov et al.}

\institute{University of Vienna, Department of Astrophysics,  Vienna, 1180, Austria; 
\email{eduard.vorobiev@univie.ac.at}
\and
Research Institute of Physics, Southern Federal University, Stachki Ave. 194, Rostov-on-Don, 
344090 Russia; 
\and
Konkoly Observatory, Research Centre for Astronomy and Earth Sciences, P.O. Box 67, H-1525 Budapest
\and
UCO/Lick Observatory, University of California, Santa Cruz, CA 95064, USA
}

\abstract 
{}
  {Motivated by recent observational and numerical studies suggesting that collapsing 
  protostellar cores may be replenished from the local environment,
  we explore the evolution of protostellar cores submerged in the external counter-rotating 
  environment. These models predict the formation
  of counter-rotating disks with a deep gap in the gas surface density separating 
  the inner disk (co-rotating with the star) and the outer counter-rotating disk. 
  The properties of these gaps are compared to those of planet-bearing gaps that form in 
  disks hosting giant planets. }   
 {We employ numerical hydrodynamics simulations of collapsing cores that are replenished from the local
 counter-rotating environment, as well as numerical hydrodynamics
 simulations of isolated disks hosting giant planets, to derive the properties of the
 gaps that form in both cases. }
  {Our numerical simulations demonstrate that counter-rotating disks can form for 
  a wide range of mass and angular momentum available in the local environment.  
  The gap that separates both disks has a depletion factor smaller than $1\%$, can be located at a distance
  from ten to over a hundred AU from the star, and can propagate inward with velocity ranging from 1~AU~Myr$^{-1}$
  to $>100$~AU~Myr$^{-1}$.
  Unlike our previous conclusion, the gap can therefore be a long-lived phenomenon, comparable in some
  cases to the lifetime of the disk itself. For a proper choice of the planetary mass, the
   viscous $\alpha$-parameter and the disk mass, the planet-bearing gaps and the gaps in 
   counter-rotating disks 
   may show a remarkable similarity in the gas density profile and depletion factor, 
   which may complicate their observational differentiation.  }
  {}
 
\authorrunning{Vorobyov et al.}
\titlerunning{Gaps in circumstellar disks}
\maketitle

\section{Introduction}


In the classic scenario for star formation, stars form from the gravitational collapse 
of dense molecular cloud cores described by
isolated Bonnor-Ebert-type spheres or truncated
singular isothermal spheres \citep{Larson1969,Shu1977}. If the core has non-zero
angular momentum, then a fraction of its mass lands onto a circumstellar disk
formed owing to conservation of the net angular momentum of the core.
In this scenario, stars and disks accrete material from isolated 
parental cores until the latter are depleted or dispersed via feedback from stellar 
irradiation and outflows.

However, there is growing evidence that forming young stellar objects 
may be constantly or periodically replenished from the external environment.
A sizeable fraction of cores appear to form not in isolation, but within dense 
and rich large-scale structures. For instance, recent Herschel and IRAM images show that dense 
cores often lie along sinuous filaments, like beads in a string \citep{Andre2014,Tafalla2015},
implying an interesting possibility for prolonged accretion from these 
filamentary structures. 

This assumption seems to be  supported by recent
analytical, numerical, and observational studies. For instance, the usually referred
mean age of 2-3~Myr for circumstellar disks \citep{Mamajek2009} might in fact be 
considerably underestimated due to selection criteria that focus onto the densest parts of stellar
clusters (most prone to disk erosion) and leave out field stars \citep{Pfalzner2014}. 
There is also a wide spread in the disk lifetimes: some objects lose their disks
very early ($\le 1.0$ Myr), whereas other objects retain their  disks for up to 10 Myr
and even longer \citep{Beccari2010,WC2011,DeMarchi2013}. 
An implication for these findings is that the disk dispersal mechanisms, 
such as UV/X-ray photoevaporation, may act longer than was previously thought 
or circumstellar disks may be replenished from the external environment.
Numerical hydrodynamics simulations of clustered star formation support the
notion of prolonged accretion, showing that protostellar cores can
be repeatedly replenished in response to the fluctuating conditions in the 
surrounding environment \citep{Bonnell2014,Padoan2014}.

Motivated by these findings, we presented recently a numerical hydrodynamics
study of collapsing cores embedded in an external environment with different
magnitude and direction of rotation \citep{VLG2015}. Our major conclusion was that 
the evolution of stars and circumstellar disks in isolated and non-isolated systems 
may be drastically different. The most curious case was found for 
the model with opposite spin directions of the core and external environment. This
peculiar system demonstrated the formation of counter-rotating disks separated by a 
deep gap in the gas surface density, resembling somewhat in morphology the AB Aurigae
system also seemingly showing the signs of prolonged accretion and counter-rotating disk
structures \citep{Tang2012}.

The formation scenario for counter-rotating disks requires a source of material, which
{\it i}) has a spin direction that is opposite (in general) to that of the disk and {\it ii})
can accrete onto disk sometime after its formation. 
The star-forming turbulent and chaotic environment may naturally provide such an environment.
Indeed, numerical simulations of clustered star formation demonstrate that 
protostellar cores can be regularly replenished in response to the fluctuating conditions 
in the local environment \citep{Bonnell2014} and the angular momentum vector of the
accreted material can undergo significant changes both in magnitude
and direction with respect to the star \citep{Bate2010, Fielding2015},
often leading to the formation of misaligned star-disk 
systems exceeding in some cases 90$^{\circ}$. The recent observations
of wide-separation ($>1000$~AU) binary/multiple systems in the Perseus
star-forming region revealed that the distribution of the outflow directions
is consistent with preferentially random or even anti-aligned distributions,
implying that these systems possibly
formed in environments where the distribution of angular momentum is complex and disordered,
rather than co-rotating or aligned \citep{Lee2016}.

\begin{table*}
\center
\caption{Model parameters}
\label{table1}
\begin{tabular}{ccccccccc}
\hline\hline
Model & $M_{\rm core}$ & $\beta_{\rm core}$ & $M_{\rm ext}$ &  $\beta_{\rm ext}$ & $\Omega_{\rm core}$  &  $r_{\rm 0}$ & $\Sigma_0$  & $R_{\rm core}$  \\
 & $M_\odot$ & $\%$ & $M_\odot$ & $\%$ & km~s$^{-1}$~pc$^{-1}$ &  AU & g~cm$^{-2}$ & pc \\
\hline
1 & 1.0 & 0.7 & 0.65 & 1.36 & 0.56  & 3430 & $3.6\times 10^{-2}$ & 0.07 \\
2 & 1.0 & 0.7 & 0.33 & 1.36 & 0.61 & 2915 & $4.3\times 10^{-2}$ & 0.068 \\
3 & 1.0 & 0.7 & 0.65 & 0.7 & 0.56 &  3430 & $3.6\times 10^{-2}$ & 0.07 \\
\hline
\end{tabular}\par
\vspace{5 pt}
{$M_{\rm core}$ and $M_{\rm ext}$ are the initial masses of the pre-stellar core and the external environment, $\beta_{\rm core}$ and $\beta_{\rm ext}$ the ratios of rotational
to gravitational energy in the core and external environment, $\Omega_{\rm core}$ the angular velocity
of the core, $r_0$ and $\Sigma_0$ the size of the central plateau and the central surface density of
the core, and $R_{\rm core}$ the radius of the core.  }
\end{table*}

Moreover, the recent numerical magnetohydrodynamics simulations of the core collapse and disk 
formation, taking into account the Hall 
effect,  suggest that the outer envelope can change its rotation direction to the opposite 
with respect to the disk if the rotation vector of the parental core and the magnetic field
are antiparallel \citep{Tsukamoto2015}. If the outer envelope remains gravitationally bound to 
the system, its subsequent infall onto the disk can serve as a source of external counter-rotating 
material.

In this paper, we perform an indepth analysis  of counter-rotating disks formed as a result
of gravitational collapse of rotating cores embedded in an external environment with the
opposite direction of rotation. We compare the properties of gaps in counter-rotating 
disks with those typically found in planet-bearing disks. The paper is organized as follows.
In Section~\ref{model}, the model description and initial conditions are provided. The
formation of counter-rotating disks is described in Section~\ref{disks} and the properties
of the gaps are provided in Section~\ref{gaps}. The comparison with the planet-bearing gaps
is performed in Section~\ref{planets} and the main conclusions are summarized in Section~\ref{summary}.

\section{Model description and initial conditions}
\label{model}
\begin{figure}
 \centering
  \resizebox{\hsize}{!}{\includegraphics{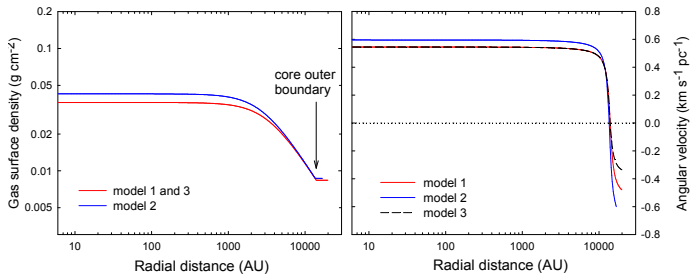}}
  \caption{Initial distribution of the gas surface density (left) and angular velocity (right) in three
  considered models.}
  \label{fig1}
\end{figure}

Our numerical model for the formation and evolution of counter-rotating disks 
is described in detail in \citet{VLG2015} and is briefly reviewed here for the reader's 
convenience. Numerical hydrodynamics simulations start from a collapsing pre-stellar core embedded into
an external low-density environment. Because the density of the core is much higher than that
of the external environment, the free-fall time of the core is short and the dynamics is 
initially dominated by contraction of the core and formation of the star plus disk system from the core
material.  

Once the mass reservoir in the core has depleted, the material from the external
environment starts falling onto the star plus disk system and the subsequent evolution 
is determined by the mass and angular momentum in the external environment. In this paper,
we consider a special case of external environment counter-rotating with respect 
to the spin of the pre-stellar core.

The main physical processes taken into account when modeling the formation and evolution of counter-rotating
disks include viscous and shock heating, irradiation by the forming star, 
background irradiation (10~K), radiative cooling from the disk surface and disk self-gravity. 
In particular, the stellar irradiation is taken into account by calculating the amount
of stellar flux intercepted by the disk/envelope. To do this, the incidence angle of 
stellar radiation is calculated from the shape of the disk surface using 
the local vertical scale height and 
the assumption of vertical hydrostatic balance. The possible self-shading
of the disk/envelope is not taken into account. The viscosity is calculated using the 
$\alpha$-parameterization of \citet{SS1973}, with the $\alpha$-value varying between 
$10^{-3}$ and $10^{-2}$. With these assumptions, the viscous and irradiation heating scale
as $r^{-3}$ and $r^{-1.75}$, respectively, so that the former usually 
dominates in the inner several tens of AU (if $\alpha$ is not too low), but the latter 
always dominates in the outer disk\footnote{ The irradiation flux is determined as 
$F_{\rm irr}=\cos \gamma_{\rm irr} L_\ast / (4 \pi r^2)$, where $L_\ast$ is the 
stellar luminosity and the cosine of the incidence angle is proportional to 
$H/r$, where $H$ is the vertical scale height. We assumed here that $H/r\propto r^{0.25}$,
but the slope in our models may vary somewhat around this value.}. 

The pertinent equations of mass, momentum, and
energy transport, and the solution procedure, are described in \citet{VLG2015}.
The forming star is described by the Lyon stellar evolution code \citep{Baraffe2012},  
while the formation and long-term evolution of the circumstellar disk 
are described using numerical hydrodynamics simulations in the thin-disk limit. 
To avoid too small time steps, we introduce a ``sink cell'' at $r_{\rm sc}=6.0$~AU and 
impose a free outflow  boundary condition so that that the matter is allowed to flow out of 
the computational domain but is prevented from flowing in.

\begin{figure*}
 \centering
  \includegraphics[width=13cm]{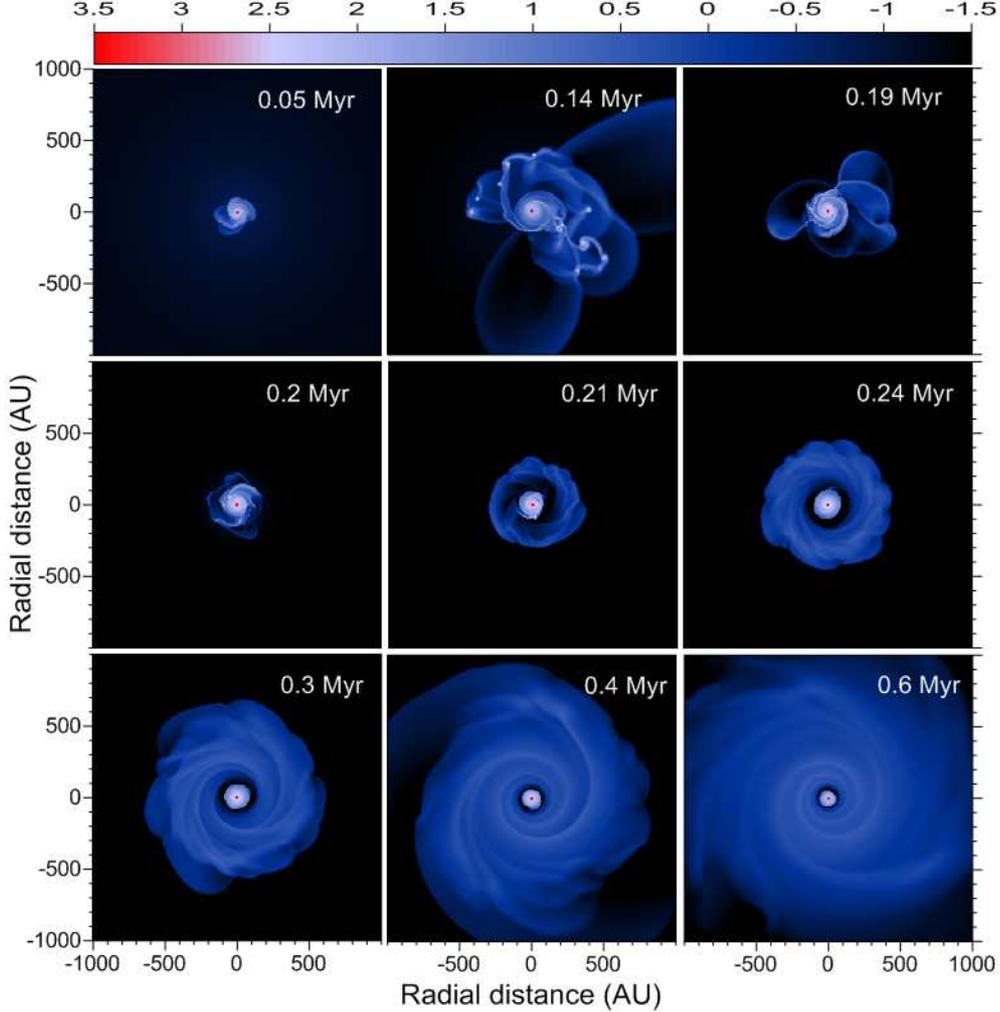}
  \caption{Gas surface density in the inner $2000\times 2000$~AU$^{2}$ box in model~1. The time elapsed since 
  the formation
  of the protostar (located in the coordinate center) is shown in each panel. The scale bar is in 
  g~cm$^{-2}$ ($\log_{10}$ units).}
  \label{fig2}
\end{figure*}

For the initial setup, we take a pre-stellar core submerged into a constant-density
external environment. For the initial surface density profile of the core, we adopt a simplified
form of a vertically integrated Bonnor-Ebert sphere \citep{Dapp09}. 
The resulting initial distribution of the gas surface density takes the following form
\begin{equation}
\label{dens}
\Sigma = \left\{
\begin{array}{ll}
{r_0 \Sigma_0 \over \sqrt{r^2+r_0^2}} \,\,\,\,\, 
       \mathrm{for}~r\le R_{\rm core}, \nonumber \\
\Sigma_{\rm ext} \hskip 0.8cm \mathrm{otherwise},            
\end{array} \right. 
\end{equation}
where $\Sigma_0$ is the gas surface
density at the center of the core, $r_0 =\sqrt{A} c_{\rm s}^2/\pi G \Sigma_0 $
is the radius of the central plateau of the core, $R_{\rm core}$ is the radius of the core, 
$c_{\rm s}$ is the initial sound speed. We assume a fixed shape of our cores with $R_{\rm core}/r_0=6$.
The radius of the core $R_{\rm core}$ is a free parameter. Once it is fixed, the size of the central
plateau $r_0$ is also fixed and $\Sigma_0$ can be found using the above expression for $r_0$, thus 
completing the procedure for generating the gas surface density distribution of individual cores.
Further, the density of the external environment  $\Sigma_{\rm ext}$ is set equal to the gas surface density at the outer edge of the core ($\Sigma_{\rm ext}=r_0 \Sigma_0/\sqrt{R_{\rm
core}^2+r_0^2}$). In all models, the value of $A$ is set to 1.2 and the initial temperature is set 
to 10~K. 

To study the formation of counter-rotating disks, we adopt the following form for
the initial radial profile of angular velocity $\Omega$ 
\begin{equation}
\label{angular}
\Omega={2\over \pi}\Omega_{\rm core} \tan^{-1} \left(  C {R_{\rm core} - r \over R_{\rm core} 
+ r } \right),
\end{equation}
where  $\Omega_{\rm core}$ is the angular velocity of the core and 
$C=25$ the dimensionless factor defining the sharpness of the transition zone between the core
and counter-rotating external environment. When $r$ is much smaller than $R_{\rm core}$, $\Omega$
approaches $\Omega_{\rm core}$.

  \begin{figure}
 \centering
  \resizebox{\hsize}{!}{\includegraphics{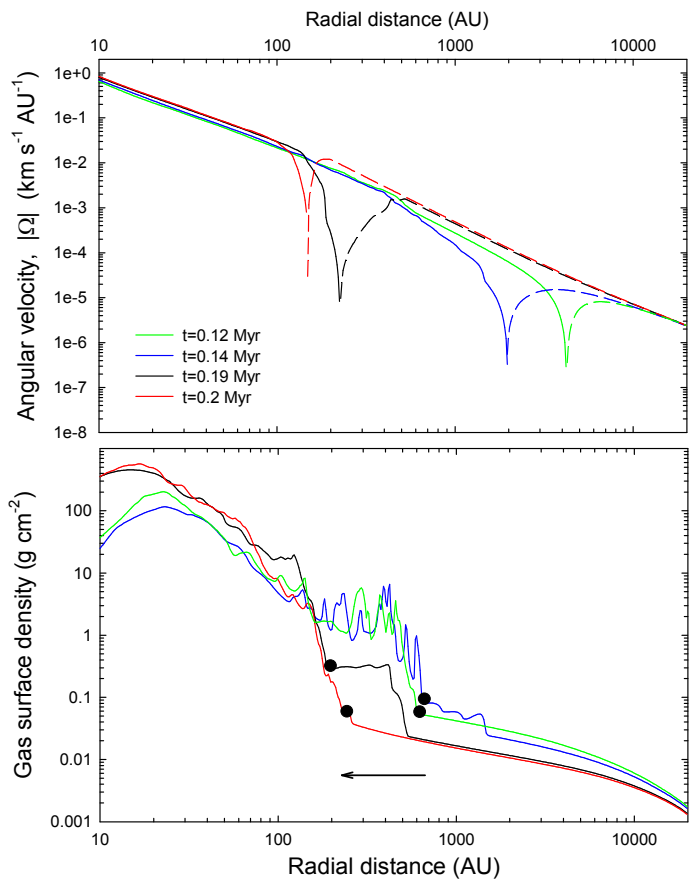}}
  \caption{{\bf Top:} Azimuthally-averaged angular velocity $|\Omega|$ (by absolute value). 
  The positive/negative values are plotted by solid/dashed lines. 
  A sharp drop in $|\Omega|$ indicates the radial position where the rotation changes its direction
  from counter-clockwise to clockwise. {\bf Bottom:} Azimuthally-averaged gas surface density. The solid
  circles mark the position of the outer disk edge. The arrow shows the direction in which the disk
  shrinks thanks to the infall of external counter-rotating material.}
  \label{fig3}
\end{figure}

We have considered three models, the initial configurations for which are shown 
in Figure~\ref{fig1}.
In particular, the left panels present the radial distribution of $\Sigma$, while the right panels
show $\Omega$ as a function of radial distance. 
In all models, the core rotates counter-clockwise, while the external environment rotates
clockwise. The parameters of every model are provided in Table~\ref{table1}.
All three models have similar integrated properties of the core, albeit with some variations in the
radial density profiles\footnote{We note that the evolution of the core depends weakly 
on the gas surface
density and angular velocity profiles, but is quite sensitive to the integrated quantities such as its
mass \citep{Vorobyov2012}}, but differ significantly in the characteristics 
of the external environment. In particular, they have different masses  $M_{\rm ext}$ and 
different ratios of rotational to gravitational energy $\beta_{\rm ext}$ in the external 
environment, while the corresponding quantities in the core 
($M_{\rm core}$ and $\beta_{\rm core}$) are the same.
In our previous study \citep{VLG2015}, the mass of the core was lower than that contained in 
the external environment. In this study, we consider the opposite initial configuration in which
the mass contained in the core is higher than that in the external environment.
We note that both the core and external environment are initially out of virial 
equilibrium ($|E_{\rm grav}| > 2 E_{\rm th} + E_{\rm rot}$) and both undergo
gravitational contraction after the start of the numerical simulations.

\section{Formation of counter-rotating disks}
\label{disks}
The formation of counter-rotating disks is demonstrated in Figure~\ref{fig2} showing 
the time evolution of the gas surface density in model~1. We zoom in onto the inner $2000\times 2000$~AU$^2$
box where the most interesting effects take place, but note that the total computational domain 
has a size of $20000\times20000$~AU$^2$. The time elapsed since the formation of the central 
protostar is shown in each panel. 

The initial configuration is gravitationally unstable and collapses to form 
a central protostar. However, because the gas density is higher in the core than in 
the external environment, the former collapses faster than the latter.
After several thousands years, a centrifugally balanced disk forms 
and grows in size and mass thanks to the continuing inflow of matter from the collapsing core. 
In this early phase, the disk corotates with the core. 
However, at $t\approx0.14$~Myr the disk growth 
halts when the mass reservoir in the collapsing core exhausts and the material from 
the collapsing counter-rotating external environment starts falling
onto the disk outer regions.  
This infalling material mixes with the disk,
reducing its net angular momentum and causing the disk to shrink by a factor of several by 
$t=0.2$~Myr. We note that the infall of external material is not modelled by adding material 
at certain radii with a given mass and angular momentum rate,
but rather through self-consistent numerical hydrodynamics simulations covering
on the same numerical mesh both the disk and the external environment 
contracting gravitationally towards the disk. 

To better illustrate the effect of infall from the external environment, we show in Figure~\ref{fig3} the angular velocity
by absolute value $|\Omega|$ (upper panel) and the gas surface density $\Sigma$ (lower panel) 
as a function of radial distance from the star. Both $|\Omega|$ and $\Sigma$ are azimuthally 
averaged. Four time instances are denoted by lines of different color as shown in the legend. 
In the upper panel, the solid/dashed lines represent positive/negative values of $\Omega$.
The inner regions rotate  counter-clockwise (positive $\Omega$), while the outer
regions rotate clockwise (negative $\Omega$). A sharp drop in $|\Omega|$ manifests a radial 
position 
where the counter-rotating external environment mixes with the inner core/disk, resulting in a
net decrease in $|\Omega|$.

\begin{figure}
 \centering
  \resizebox{\hsize}{!}{\includegraphics{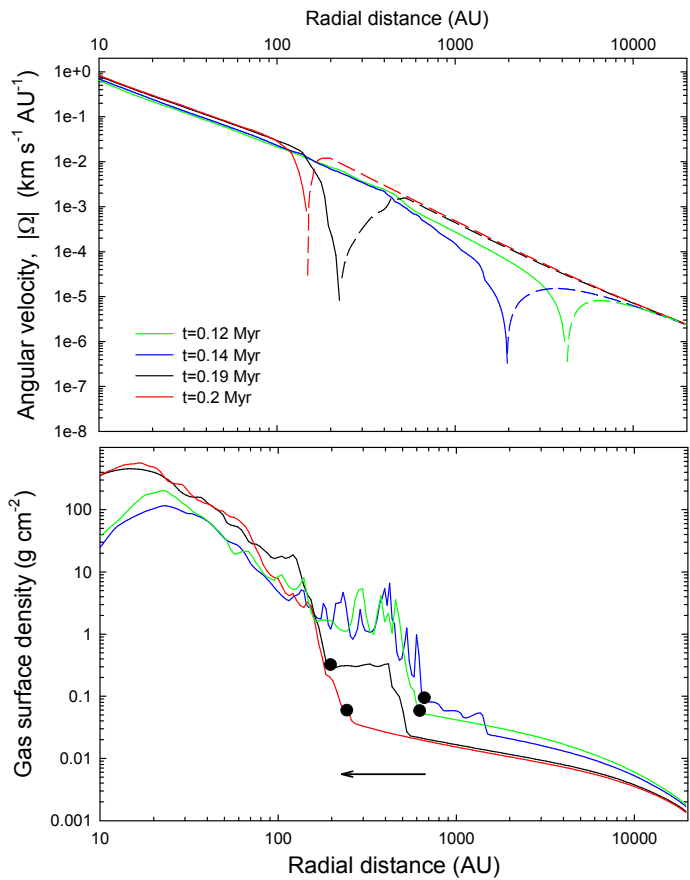}}
  \caption{Gas velocity field superimposed on the gas surface density map 
  in model~1 at $t=0.24$~Myr. Two counter-rotating disks are clearly visible.}
  \label{fig4}
\end{figure}

The upper panel in Figure~\ref{fig3} demonstrates that the interface between 
the external environment (dashed lines)
and the inner material constituting the core and the disk (solid lines) moves radially 
inward with time. At $t\le0.14$~Myr,
the position of the interface is still far away from the disk outer edge, the latter 
is schematically indicated in the lower panel by the filled circles. 
As a result, the disk grows in size during the early evolution thanks to
continuing accretion of corotating material from the collapsing core. However, once 
the core has almost completely accreted onto the disk plus star system, the 
material from the counter-rotating external environment starts landing onto and 
mixing with the disk. The  
latter shrinks in size as indicated by the arrow in the lower panel. 
During $t=0.19-0.20$~Myr, the size of the disk is determined by the current position 
of the interface between the disk and the infalling external material.

The subsequent evolution of the system ($t\ge0.21$~Myr) reveals an interesting effect: the 
formation of two counter-rotating disks separated by a deep gap. This phenomenon, first noted
in \citet{VLG2015}, is 
illustrated in Figure~\ref{fig4} showing the gas velocity field (yellow arrows) superimposed
on the gas surface density map at $t=0.24$~Myr.  The inner disk rotates counter-clockwise, in the 
same direction as the initial rotation of the parental core, while the outer disk rotates clockwise, in 
the same direction as the external environment. As time passes, the outer disk gradually grows in 
size owing to infall of the external material, while the inner disk shrinks owing to accretion 
onto the star. 

\begin{figure}
 \centering
  \resizebox{\hsize}{!}{\includegraphics{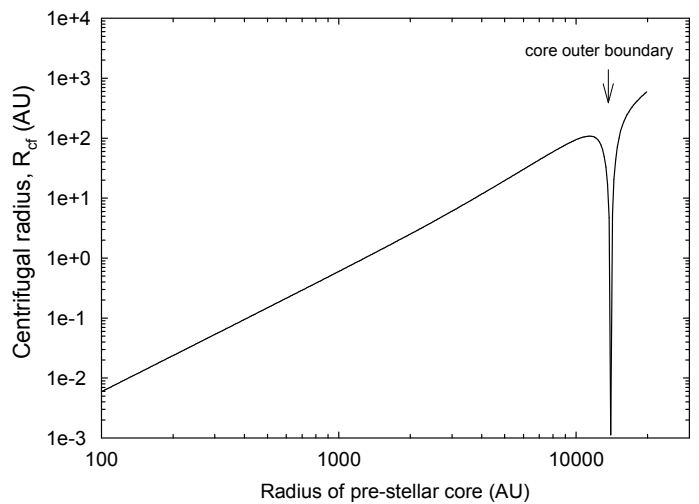}}
  \caption{Centrifugal radius $R_{\rm cf}$ as a function of radial distance in model~1
  at the onset of gravitational contraction (t=0). The arrow
  indicates the position of the core outer boundary.}
  \label{fig5}
\end{figure}

The formation of the outer counter-rotating disk can be understood if we consider 
the centrifugal radius of matter initially located at a distance $r$ from the star\footnote{
We note that this equation is strictly valid for a point-mass object. However, 
our initial surface density distribution is similar to that of Mestel's disk, 
$\Sigma \propto r^{-1}$, the gravitational potential of which is similar to that of the point-mass object
with the point mass substituted by the mass 
located within a given radius $r$ \citep{Binney}. That is why Equation~(\ref{centrad}) 
is also applicable to our models. }:
\begin{equation}
R_{\rm cf}= {J^2(r) \over G M(r)},
\label{centrad}
\end{equation} 
where $J(r)=r^2 |\Omega|$ is the specific angular momentum at a radial distance $r$,
$G$ is the gravitational constant, and $M(r)$ it the mass enclosed within distance $r$.
The centrifugal radius provides a distance at which the gravitational acceleration 
acting on a contracting layer of rotating material becomes balanced by its centrifugal acceleration,
preventing the layer from further contraction.
The radial distribution of $R_{\rm cf}$ in model~1  at the onset of gravitational
contraction ($t=0$) is shown in Figure~\ref{fig5}.
The centrifugal radius gradually increases with distance and reaches a local maximum 
at the location of the core outer edge (shown by the arrow). In the transition zone between the 
core and the counter-rotating external environment, the angular velocity changes its sign and
the centrifugal radius drops to a negligible value. 
In the counter-rotating external environment, the angular velocity increases again, 
though with the opposite sign. 
As a consequence, the centrifugal radius of the external environment increases and 
attains values that are even larger than those of the core.

The disk evolution reflects the behaviour of the centrifugal radius 
shown in Figure~\ref{fig5}. First,
the disk grows in size owing to accretion of material with a gradually increasing $R_{\rm cf}$ (the
value of which serves as a proxy for the disk radius).
The centrifugal radius of the material at the core outer edge is $\approx 200$~AU. We note that
the disk at this stage is somewhat larger
owing to gravitational interaction between spiral arms and fragments, which leads
to gravitational scattering of the fragments and effective increase of the disk radius.
The growth of the disk is followed by tentative contraction when the low-$R_{\rm cf}$ material 
from the transition zone lands onto the disk and starts extracting disk's angular momentum.
Finally, the outer counter-rotating disk begins to form when the infalling external material 
characterized by
large $R_{\rm cf}$ hits the centrifugal barrier just outside the heavily reduced inner disk.
The transition region between the inner and outer disks, where rotation changes its direction and
the matter lacks centrifugal support, is manifested by a density gap clearly visible 
in Figure~\ref{fig2} at $t\ge0.21$~Myr.

\begin{figure*}
 \centering
  \includegraphics[width=13cm]{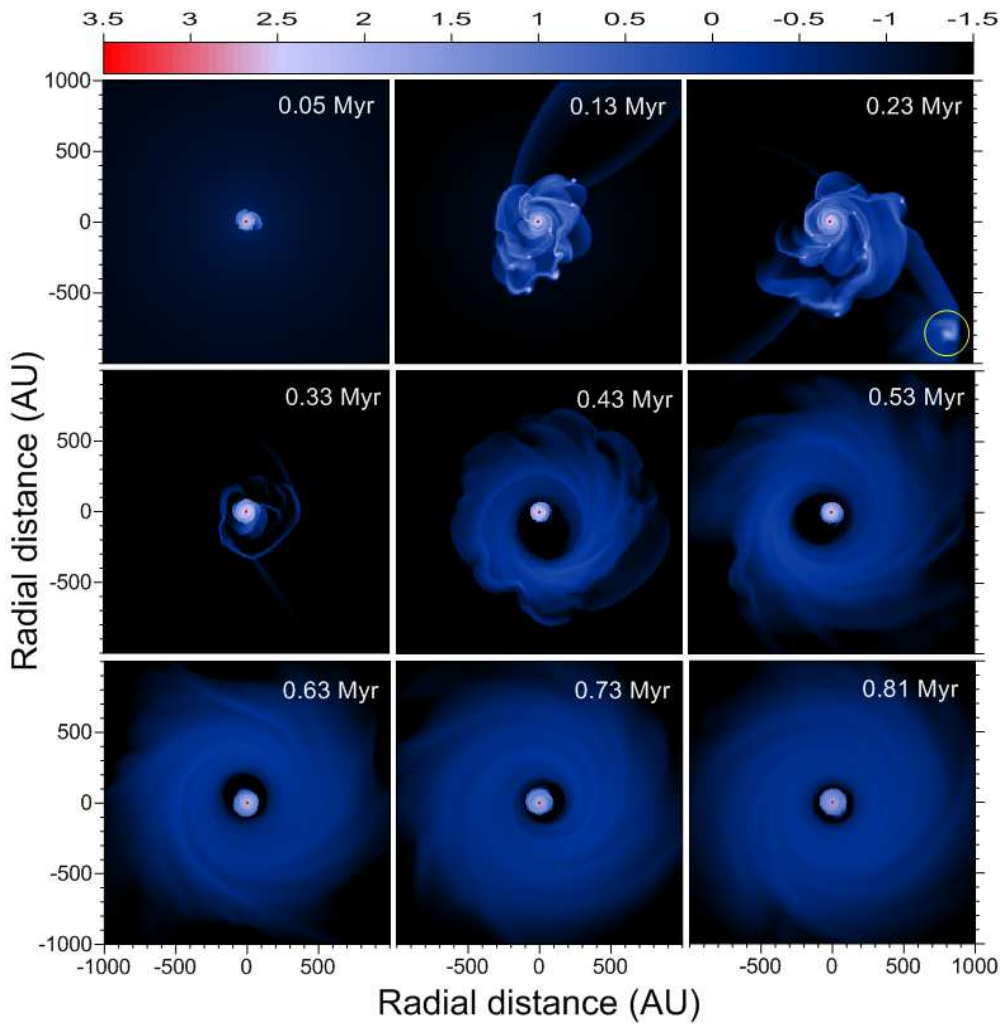}
  \caption{Similar to Figure~\ref{fig2} only for model~2. The yellow circle outlines the fragment
  ejected from the inner disk.}
  \label{fig6}
\end{figure*}

Now, we briefly discuss the formation of counter-rotating disks separated by a gap in the 
other two models considered in our work.
Figure~\ref{fig6} demonstrates the formation of the counter-rotating disks in model~2, which is characterized
by a somewhat smaller available mass reservoir in the external environment than model~1 (see Table~\ref{table1}).
The time elapsed since the formation of the star (located in the coordinate center) is shown in each
panel. The evolution in model~2 follows a path similar to that of model~1: the initial growth of 
the disk is followed by its contraction and formation of a counter-rotating outer disk. However, 
the outer disk and the gap are initially much more eccentric than in model~1.

To illustrate this peculiar feature of model~2, we calculated the eccentricity as 

\begin{equation}
        \label{eq:ecc}
        e(r,\phi)=\sqrt{1+2h(r,\phi) c(r,\phi)^2}.
\end{equation}

In Equation~(\ref{eq:ecc}), $c(r,\phi)$ and $h(r,\phi)$ stand for
\begin{equation}
        \label{eq:c}
        c(r,\phi)=x(r,\phi)v_\mathrm{y}(r,\phi)-y(r,\phi)v_\mathrm{x}(r,\phi),
\end{equation}
and
\begin{equation}
        \label{eq:h}
        h(r,\phi)=\frac{v_\mathrm{x}(r,\phi)^2+v_\mathrm{y}(r,\phi)^2}{2}-\frac{1}{\sqrt{x(r,\phi)^2+y(r,\phi)^2}},
\end{equation}
where $v_\mathrm{x}(r,\phi)$, $v_\mathrm{y}(r,\phi)$ and $x(r,\phi)$, $y(r,\phi)$ are the Cartesian velocity components and coordinates at the polar grid with coordinates $(r,\phi)$.

\begin{figure}
 \centering
  \resizebox{\hsize}{!}{\includegraphics{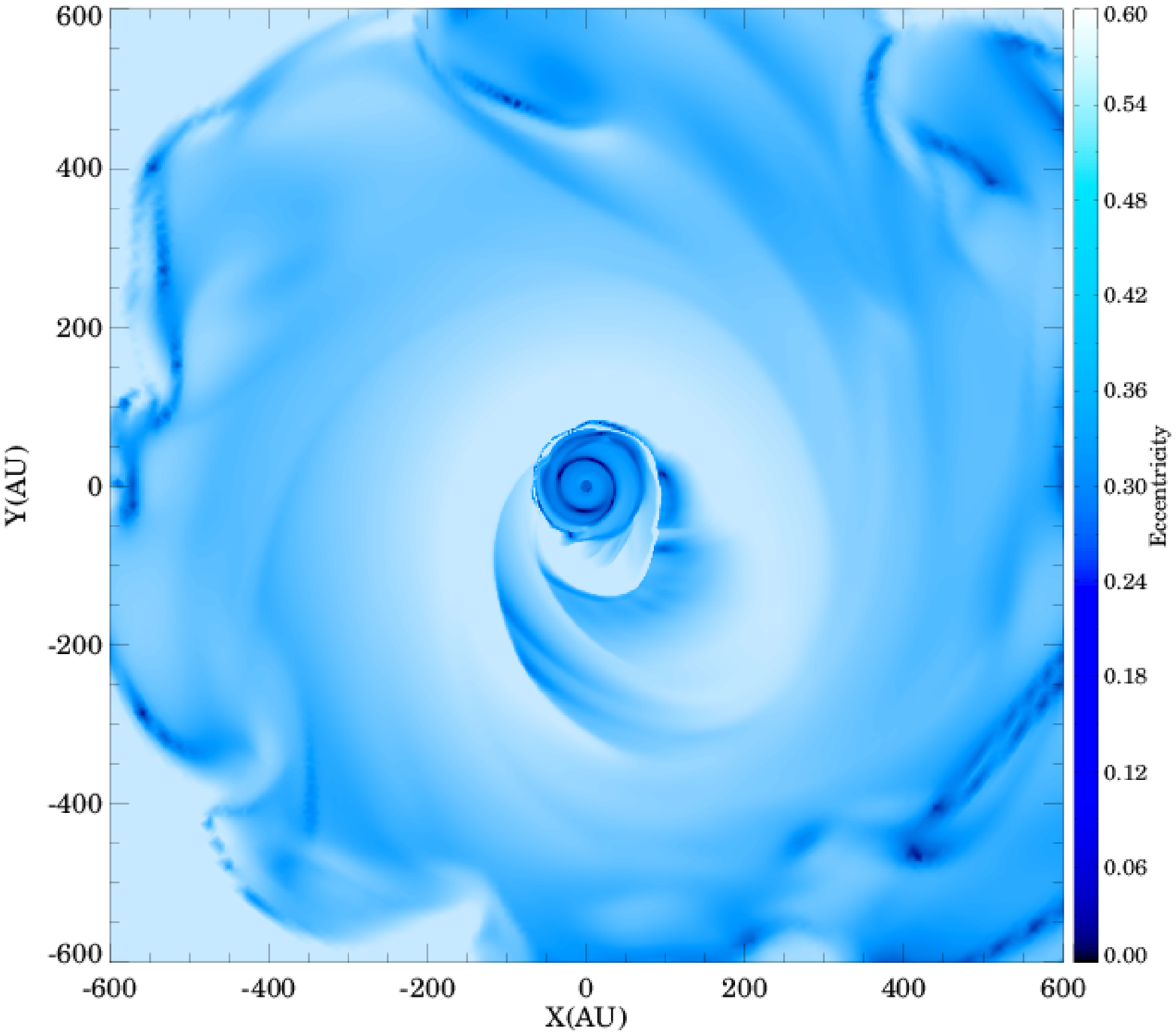}}
  \resizebox{\hsize}{!}{\includegraphics{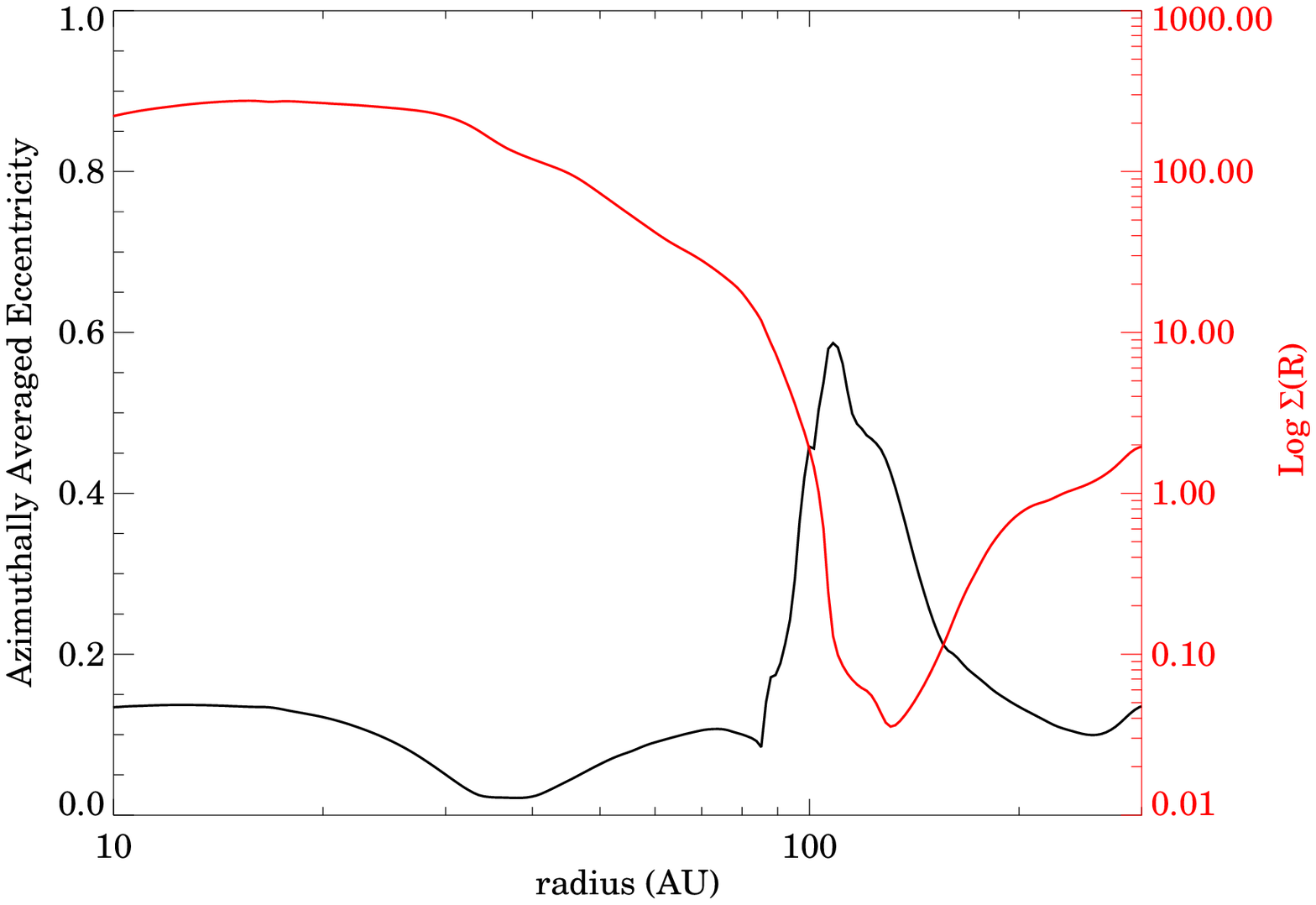}}
  \caption{Eccentricity distribution in the disk (upper panel) and the azimuthally averaged eccentricity (lower panel) for model 2 calculated using Eqs. (\ref{eq:ecc}), (\ref{eq:c}), and (\ref{eq:h}) at 0.24\,Myr.
The red line in the lower panel provides the azimuthally averaged gas surface density profile.
}
  \label{fig7}
\end{figure}

Figure~\ref{fig7} presents the map and azimuthally averaged profile of the eccentricity in 
model~2 for the inner $400\times 400$~AU$^2$ box. The snapshot is taken at 
$t=0.53$\,Myr. Evidently, 
the eccentricity is excited in the vicinity of the gap edges with a peak value of $e_{\rm peak}\approx0.6$ at the gap inner wall. We emphasise that the disk eccentricity profile is similar to that of a 
giant-planet-bearing disk (i.e., the eccentricity has a  peak inside the gap), although the latter 
have a somewhat lower amplitude at the peak value, $e_{\rm peak}\approx0.3$  \citep{Regaly2010}. 
Due to the 
high disk eccentricity, we expect strong kinematical signatures similar to those of young
binaries predicted by \citet{Regaly2011}.

 We propose the following explanation for the formation of such an eccentric outer disk. 
The initial distribution
of gas in the external environment is homogeneous and axisymmetric, so the initial conditions cannot
be responsible for the resulting eccentricity.  We think that high 
eccentricity of the gap and its edges is caused by the gravitational perturbation from the 
strongly asymmetric structure of  the inner disk. Indeed, the early evolution of the inner disk is characterized
by the presence of extended asymmetric spiral arcs. There is also a massive fragment 
($\approx25-30~M_{\rm Jup}$) outlined by the yellow circle 
in the bottom-right corner of Figure~\ref{fig6} at $t=0.23$~Myr, which is likely ejected 
from the disk via multi-body interaction with other fragments. 
These strong azimuthal distortions may cause
strong perturbations to the infalling external material, which results in the development of the 
eccentric gap and outer disk. We note that the inner disks in model~1 and model~3 (see below) 
also exhibit a spiral structure and fragmentation before the formation of the outer disk, 
but nevertheless we do not see strong eccentricity developing in the outer disk during the evolution. It might be due to the fact that models~1 and 3 have a factor of 2 greater mass reservoir 
in the external environment and, as a consequence more massive outer disks, which are less prone to
the eccentricity excitation than that of model~2.
In any case, it appears that the outcome depends significantly on the available mass in the external
environment, on particular
inner disk configuration including the presence or absence of extended spiral arcs and ejected fragments,
and therefore cannot be predicted a priori.

Figure~\ref{fig6} indicates that the eccentricity of the disk and the gap in model~2
seems to diminish with time. This is likely due to the absence of continuing gravitational perturbation
caused by smooth density distribution of the inner disk. Indeed, the distribution 
of gas in the inner disk
(after the formation of the outer disk) becomes nearly axisymmetric, likely due to disk contraction
and associated heating, thus reducing the source of gravitational
perturbation. As a result, the disk eccentricity is gradually damped by the viscous evolution.

\begin{figure*}
 \centering
  \includegraphics[width=13cm]{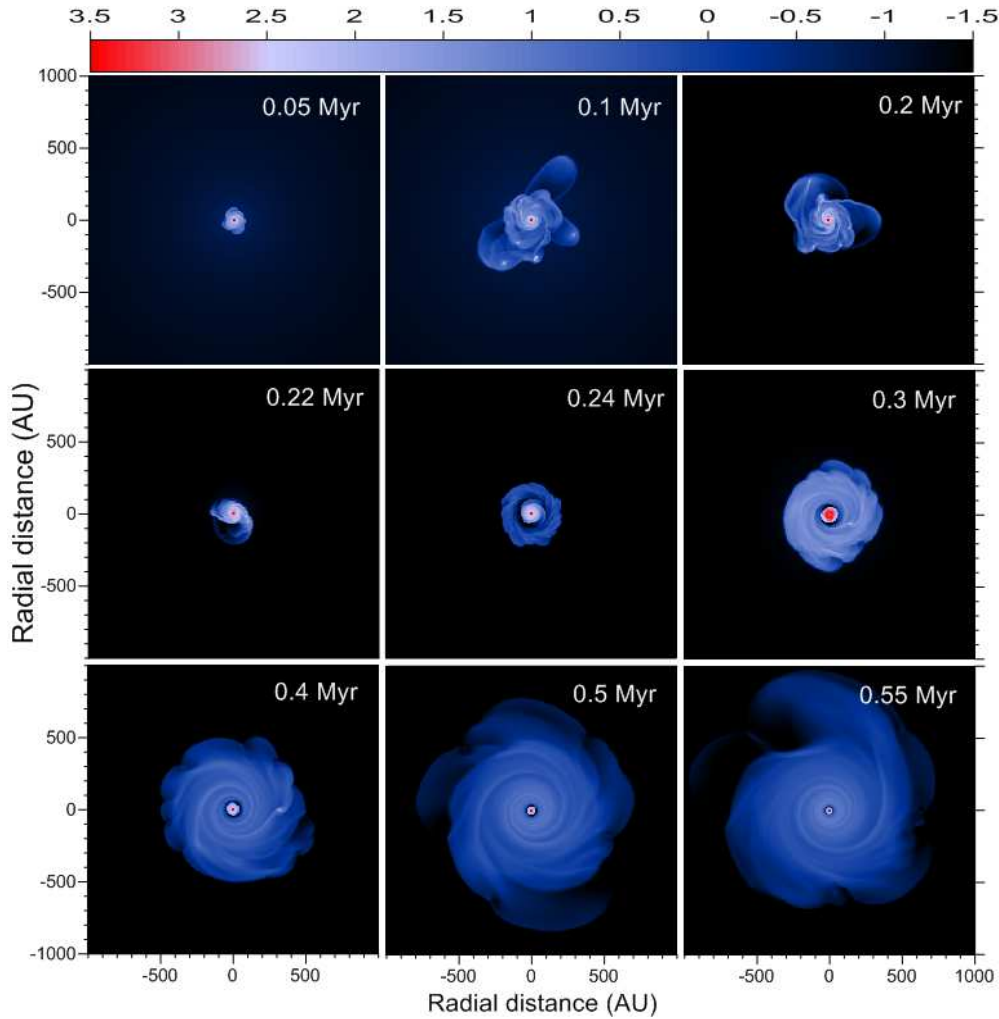}
  \caption{Similar to Figure~\ref{fig2} only for model~3.}
  \label{fig8}
\end{figure*}

Finally, in Figure~\ref{fig8} we show the formation of counter-rotating disks in model~3. 
This model is characterized by a factor of 2 smaller value of $\beta$ in the external environment
than in model~1. Nevertheless, the overall evolution is similar to model~1, except that the gap
appears to be somewhat narrower. We conclude that the gap can form for a wide range of physical 
parameters (masses, angular momenta) in the external environment.
Another interesting phenomenon that can be seen in Figure~\ref{fig8} is gravitational
fragmentation in the outer counter-rotating disk at $t=0.3-0.4$~Myr. A similar effect was also reported
in \citet{VLG2015}. Although the fragment does not survive for long (which might be caused by insufficient
numerical resolution), this phenomenon present an interesting gateway for the formation of giant planets
counter-rotating with respect to the host star.  

\begin{figure}
 \centering
  \resizebox{\hsize}{!}{\includegraphics{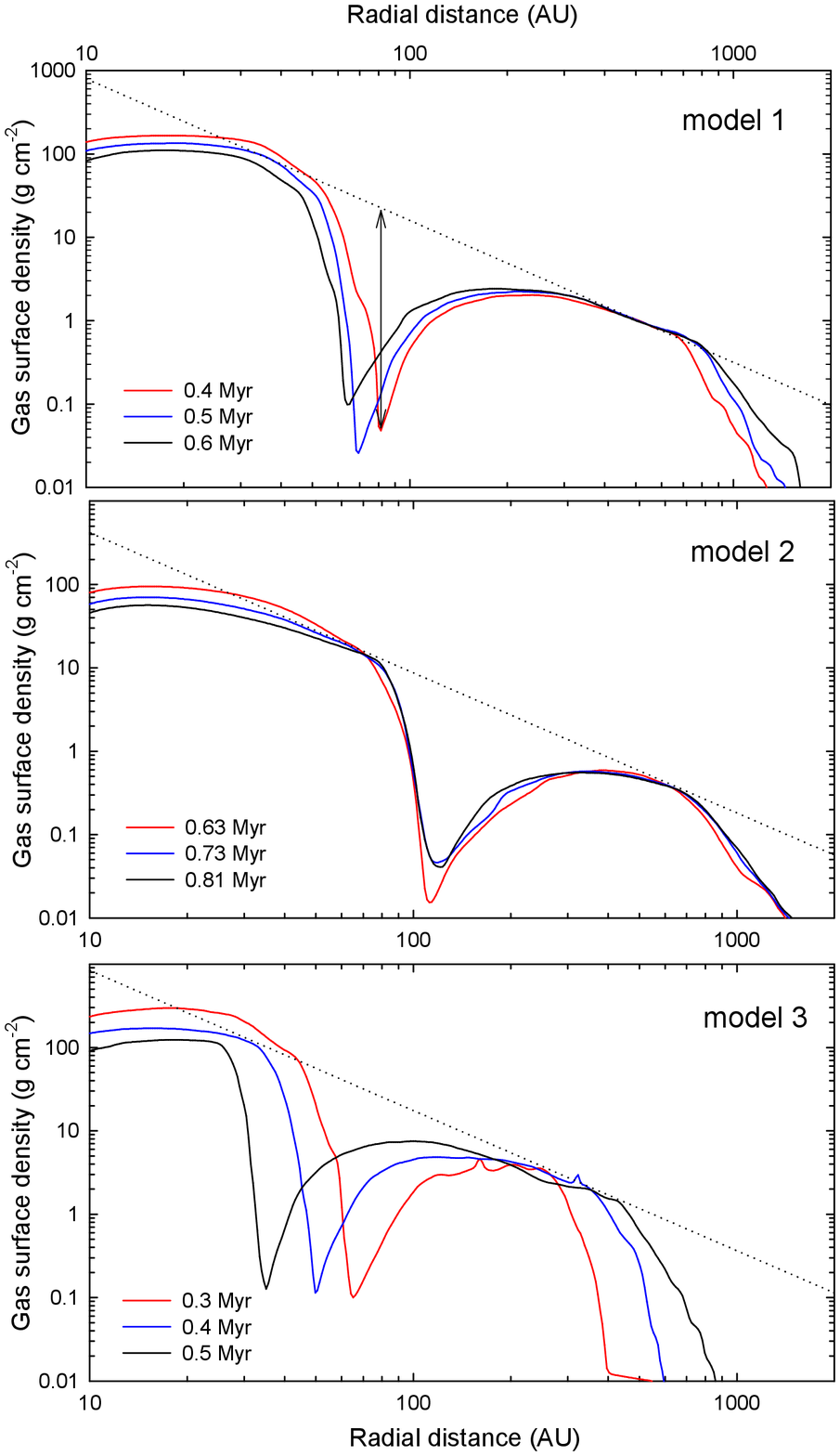}}
  \caption{Azimuthally averaged profiles of the gas surface density in models 1, 2, and 3 (from top
  to bottom). Only late evolution times, when the gap has fully developed, are shown. The dotted tangent
  lines connect the radial density profiles in the inner and outer disks. The arrow illustrates
  the contrast in the gas surface density in the gap at t=0.3~Myr. The time is counted from the formation
  of the central star.}
  \label{fig9}
\end{figure}

\section{Properties of the gaps}
\label{gaps}
In this section we consider the properties of the gaps formed in our models. Figure~\ref{fig9}
presents the azimuthally averaged profiles of the gas surface density in three models. Only the late
evolution times are considered. A deep gap in the gas surface density
is evident between the inner and outer disks. In order to estimate the depth of the gap, 
we plot a tangent line connecting the radial density profiles in the inner and outer disks 
shown with the dotted lines. The depletion of the gap is then calculated as the ratio between 
the gas surface density given by the model and that given by the tangent, 
both calculated at the position of the gap's deepest point.
The arrow illustrates the gap depletion for model~1 at $t=0.4$~Myr. 
Table~\ref{table2} provides the time-averaged properties of the gaps.

\begin{table}
\center
\caption{Properties of the gaps}
\label{table2}
\begin{tabular}{cccc}
\hline\hline
Model & $R_{\rm gap}$ (AU) & $\triangle \Sigma_{\rm gap} (\%)$ & $u_{\rm gap}$ (AU~Myr$^{-1}$)  \\
\hline
1 & 72 & 0.14 & 83 \\
2 & 115 & 0.39  & $<1.0$  \\
3 & 50 & 0.18 & 150  \\
\hline
\end{tabular}\par
\vspace{5 pt}
{ ${R}_{\rm gap}$ is the
mean position of the gap, $\triangle \Sigma_{\rm gap}$ the gap depletion in per cent, 
and $u_{\rm gap }$ the mean propagating velocity. }

\end{table}

Evidently, the gaps are characterized by a wide range of radial positions, which 
may vary from a few tens of AU to more than a hundred AU. In particular, model~2 is
characterized by the gap position that is roughly a factor of two further out than
in the other two models. This is related to the fact that model~2 has the
smallest external mass reservoir. The depletion factor $\triangle \Sigma_{\rm
gap}$ in all models is rather strong, indicating a drop in the gas surface density
by at least two orders of magnitude at the position of the gap. 
The strongest contrast 
between the models is found for the gap propagation velocity, which may vary from rather fast 
propagation in model~3 (AU~Myr$^{-1}$) to very slow inward motion in model~2 ($<1$~AU~Myr$^{-1}$). 
Model~3 is characterized by the smallest angular momentum in the external environment ($\beta$=0.7).
In our previous work \citep{VLG2015}, we showed that infall of external material with low angular
momentum exerts a strong negative torque onto the inner disk, which can lead to significant 
contraction or even complete dissipation of the latter. A similar effect takes place in model~3,
promoting the contraction of the inner disk and increasing the gap propagation speed.
Interestingly, model~2 with the slowest gap propagation velocity is also characterized
by the greatest asymmetry in both the gap and the outer disk shape. 
This implies a causal link between the shape of the gap and its propagation velocity.

In \citet{VLG2015} it was claimed that the gaps formed in counter-rotating disks are short-lived (several
tens of kyr), transient phenomena. This conclusion was based on the behaviour of the gap in 
one model only. Our numerical simulations with a wider parameter space show that the gaps may 
be a long-lived phenomenon, lasting for at least several hundred of kyr and perhaps even longer.

Finally, we note that the slopes of the tangents in Figure~\ref{fig9} lie in the $\alpha_{\rm
d}=-[1.6:1.8]$
range, which is only slightly steeper than expected for young self-gravitating disks, 
$\alpha_{\rm d}=-1.5$ \citep{Vor2010,Rice2010}, and
significantly steeper than expected for more evolved, viscosity-dominated protoplanetary 
disks with the kinematic viscosity proportional to the disk radius, $\alpha_{\rm d}=-1.0$ 
\citep{LBP1974,WC2011}.

\section{Comparison with planet-bearing gaps}
\label{planets}

\begin{figure*}
\includegraphics[width=2\columnwidth]{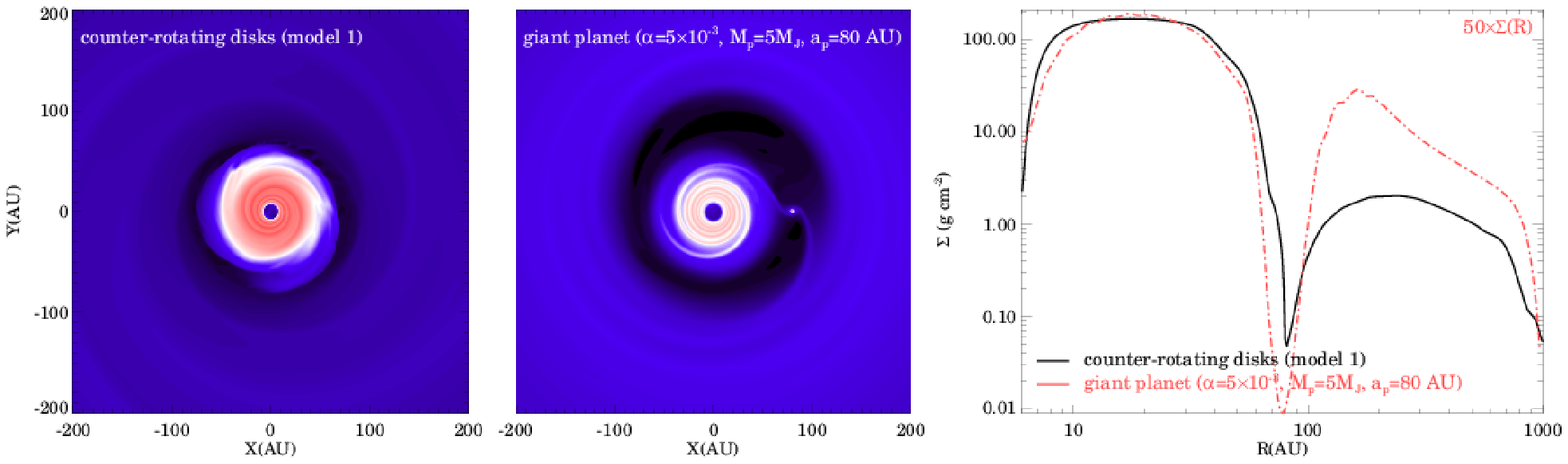}
\includegraphics[width=2\columnwidth]{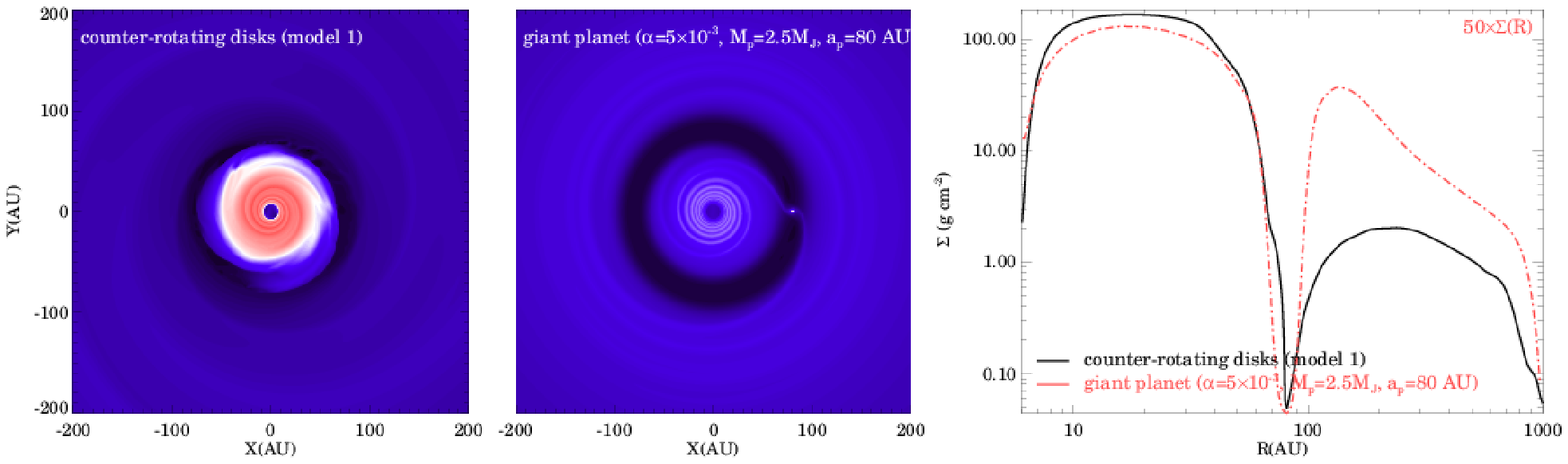}
\caption{Comparison of gap formation in counter-rotating disks and a giant plant bearing protoplanetary 
disk assuming $5\,M_\mathrm{Jup}$ (upper row) and $2.5\,M_\mathrm{Jup}$ (lower row) planetary mass, respectively. 
\emph{Left panel:} density distribution in the counter-rotating disk models at 0.6\,Myr. \emph{Middle panel}: 
density distribution in giant planet bearing disk model after 1000 orbits of the giant planet. \emph{Right panel:} 
comparison of azimuthally averaged density profiles for both models. The density is scaled 
up by 50 in planet-bearing disk model to mach the density profiles in the inner disk.}
\label{fig10}
\end{figure*}

\begin{figure*}
\includegraphics[width=2\columnwidth]{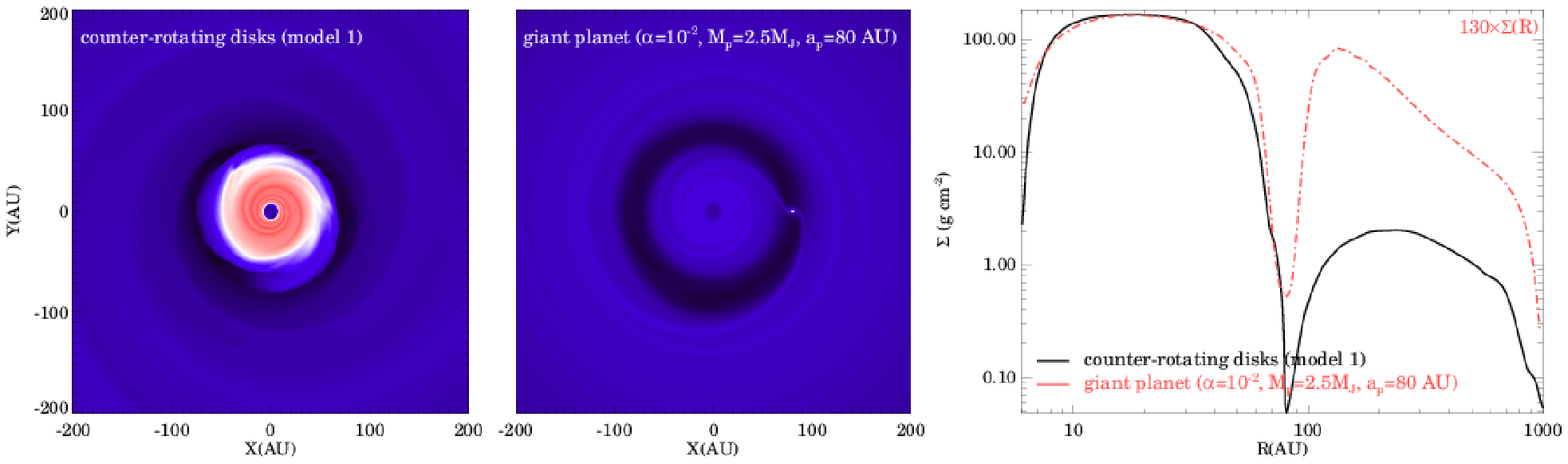}
\includegraphics[width=2\columnwidth]{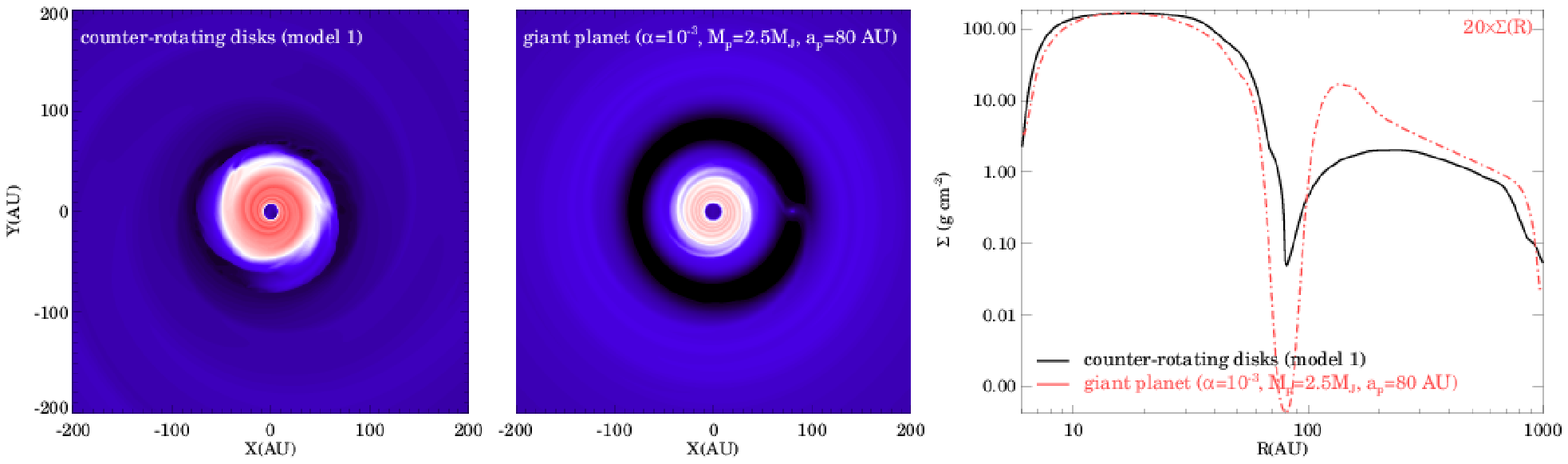}
\caption{Same as Fig.\,\ref{fig10}, but the viscosity is changed to $\alpha=10^{-2}$ (upper row) 
and $\alpha=10^{-3}$ (lower row). The applied density scalings are 130 (upper row) and 20 (lower row).}
\label{fig11}
\end{figure*}

\begin{figure*}
\includegraphics[width=2\columnwidth]{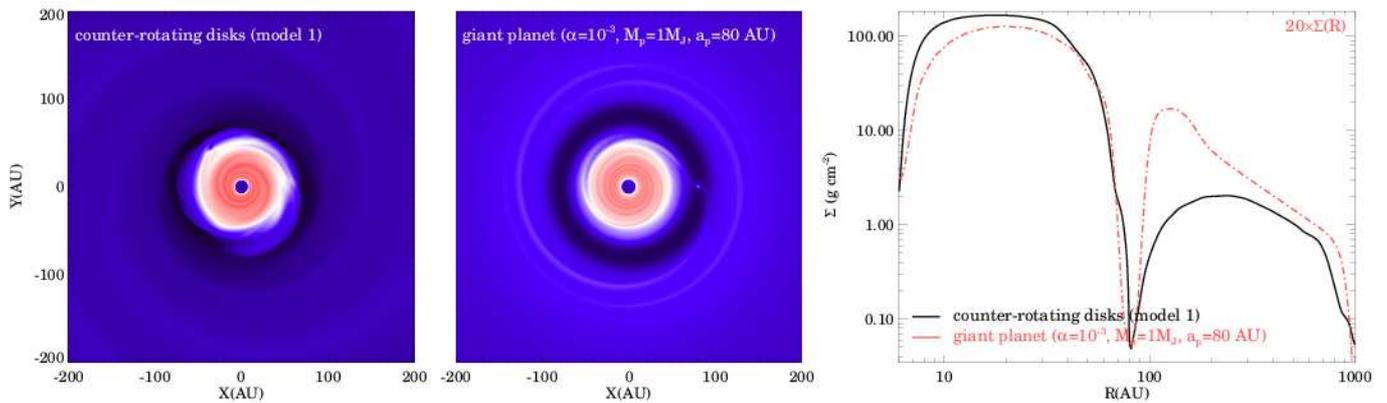}
\caption{Same as the lower panel  of Fig.\,\ref{fig11} ($\alpha=10^{-3}$), but the planetary mass is 
decreased further to $1M_\mathrm{Jup}$. The applied density scaling is 20.}
\label{fig12}
\end{figure*}

\begin{figure*}
\includegraphics[width=2\columnwidth]{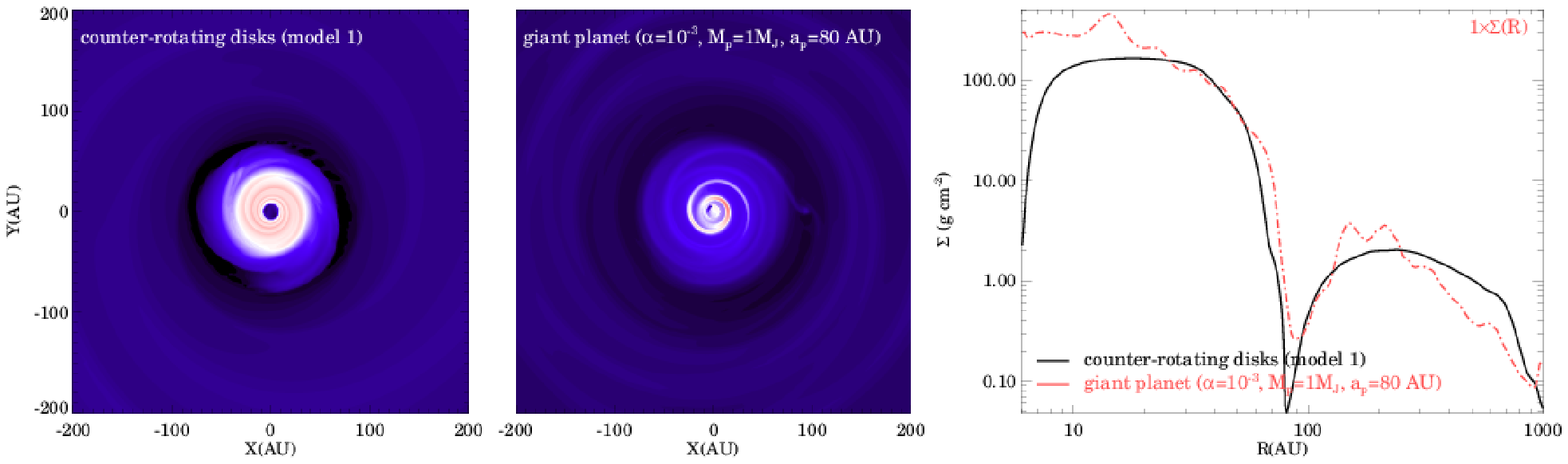}
\caption{Same as in Fig.\,\ref{fig12}, but in models where the disk mass is assumed to 
be 10 times higher, therefore the self-gravity is taken into account. No density scaling is 
applied.}
\label{fig13}
\end{figure*}

In this section, we compare the properties of gaps formed in counter-rotating disks with those 
that can be found in disks with embedded planets. A massive planet clears the disk around it to form
a gap through gravitational interaction if its Hill radius is greater than the local disk scale height
and the time scale for the gap opening is shorter than the time scale for viscous diffusion to close
it \citep{KN2012}. To model the gap opening via planet-disk interaction, we employ 
the two-dimensional numerical 
hydrodynamics simulations using the FARGO-ADSG code and also its GPU based version (Masset 2000). 
The FARGO code solves the vertically integrated continuity and Navier--Stokes equations 
numerically in the cylindrical coordinate system, using the locally isothermal approximation.

In the local isothermal approximation, the radial temperature profile is $T(r) \sim r^{-1}$, and 
the equation of state of the gas is ${\cal P}(r,\phi)=c_\mathrm{s}(r)^2\Sigma(r,\phi)$ depends 
only on the density and the local sound speed $c_\mathrm{s}(r)=\Omega(r)H(r)$, where $H(r)$ 
is the local pressure scale 
height. Note that both $c_\mathrm{s}(r)$ and $H(r)$ are constant in time due to the assumption
of local isothermality. The disk scale height is proportional to radius, $H(r)=A r$, where 
$A$ is the disk aspect ratio, which is set to $A=0.05$ for all models. 

The 2D computational domain extends from 6~AU to 1000~AU, consisting of $N_R=256$ logarithmically 
distributed radial, and $N_\phi=512$ equidistant azimuthal grid cells. For the smoothing length
of the planet gravitational potential we choose $\epsilon H(R_\mathrm{p})$, where
$R_{\rm p}$ is the position of the planet and $\epsilon=0.6$, the latter being appropriate for 
2D simulations \citep{Kley2012}. The gas is allowed to flow out from the 
computational domain, i.e. an open boundary condition is applied at both the inner 
and outer domain boundaries. The disk viscosity is approximated by the $\alpha$-prescription of 
\citet{SS1973}. The magnitude of viscosity is varied in the range of 
$10^{-3}\leq\alpha\leq10^{-2}$. For comparison, in the counter-rotating disk simulations, 
the viscous $\alpha$ was set to a constant value of $5\times 10^{-3}$.
We use a coordinate frame that co-rotates with the planet.

The initial density profile of the gas disk is $\Sigma(R,\phi)=\Sigma_0 R^{-1.5}$, where 
$\Sigma_0$ is chosen so as to set the disk mass to $\sim0.03~M_\odot$ inside the computational 
domain. For this disk mass, the  Toomre $Q$-parameter \citep{Toomre1964} is greater than three 
throughout the disk ($Q\simeq8.5$ at the planetary distance). These high values of $Q$-parameter 
make it possible to neglect the disk's 
self-gravity. We also consider models for which the disk mass is increased to $\sim0.3~M_\odot$. 
For those high-disk-mass models, the disk's self-gravity is taken into account.

The planet is set on an orbit at $80$\,AU from the central star (to match the position of the
gap in the counter-rotating disk model) and the planet is not allowed to migrate or accrete 
gas, i.e., it remains on a circular orbit with constant mass. The planet-to-star mass ratio are 
$q=0.9\times10^{-3}$, $2.35\times10^{-3}$, and $4.7\times10^{-3}$ corresponding to 
$1,\,2.5,\,$ and $5$ Jupiter mass planets for a $1.0$ Solar mass central star. The planetary 
mass and the disk viscosity were 
varied in a wide range to produce different configuration of the planet-bearing gaps. 
Our purpose is to determine the characteristic signatures of the gaps in counter-rotating 
disks which can help to distinguish them from the planet-bearing gaps. 

For a planet-bearing disk model, the density snapshots were taken at the end of the 
simulation (after 1000 orbits of the giant planet corresponding to 0.07 Myr after the planet 
was inserted to the disk), when a quasi 
steady disk state had been achieved. In each row of Figures~\ref{fig10}-\ref{fig13}, the first 
and second panels show the gas surface density distribution in the counter-rotating disk 
and planet-bearing disk models. The third panel shows the azimuthally averaged radial density 
profiles (black for counter-rotating, blue for planet-bearing disk models). 
For the counter-rotating disk model, we use the data for model~1 at $t=0.6$~Myr.
The densities in non-self-gravitating planet-bearing models are scaled such that 
the density profile at the inner disk ($R<80$\,AU) fits to that of the counter-rotating 
disk model. We note that as long as the disk self-gravity and the planetary migration are 
not taken into account, the density can be scaled as $\Sigma_0$ can 
be cancelled out from the continuity and Navier--Stokes equations.

Figure\,\ref{fig10} shows the comparison of gaps in counter-rotating and planet-bearing 
disk models for the viscous $\alpha=5\times10^{-3}$ and the planet mass of $M_{\rm p}=5$ 
(upper row) and $M_{\rm p}=2.5$ (lower 
row) Jupiters. The surface density in the planet-bearing model is scaled up by a 
factor of 50 in order to match 
the peak density in the inner disk of the counter-rotating disk model\footnote{We scale 
$\Sigma$ with a sole purpose of facilitating the comparison between the radial density profiles
in different models. This scaling procedure is not meant for comparing the absolute values 
of the gas surface density, but only their ratios, such as the gap depletion factor.}. 
We found that the lower 
$M_{\rm p}=2.5$~$M_{\rm Jup}$ model seems to match slightly better the azimuthally averaged 
density profiles in the inner 
disk, including the density contrast between the inner disk edge and the center of the gap, 
whereas both planet-bearing models are characterized by a smaller contrast between the peak
densities in the inner and outer disks. 
As a consequence, the depletion in planet-bearing gaps is significantly stronger
than in the case of counter-rotating disks, 0.02\% 
for $5\,M_\mathrm{Jup}$ and 0.08\% for $2.5\,M_\mathrm{Jup}$ planet as compared to
0.14\% in model~1.

Figure\,\ref{fig11} shows the effect of viscosity on the gas density distribution in 
planet-bearing disk models with different values of $\alpha=10^{-2}$ (upper row) and $\alpha=10^{-3}$ (lower row). 
The planet mass in both cases is set to $M_{\rm p}=2.5$~$M_{\rm Jup}$. 
For both models, we applied a density scaling factor of 130 and 20 to match the
inner density profile of the counter-rotating disk model. The model with a higher 
$\alpha=10^{-2}$ demonstrates a significantly shallower gap (the depletion factor is 0.37\%) 
than its low-$\alpha$ counterpart (0.002\%) thanks to an increased viscous transport. 
As a consequence, the gap depth in the $\alpha=10^{-2}$ model fits better
to that of the counter-rotating disk model.

To better match the density depletion inside the gap, the planetary mass was further decreased 
to $1.0~M_\mathrm{Jup}$. The corresponding density distribution is shown in Figure\,\ref{fig12}. 
For this case, the proper density scaling for the planet-bearing model was found to be 20. 
With $1.0~M_\mathrm{Jup}$ for the planetary mass and $10^{-3}$ for the viscous $\alpha$, 
the inner disk density profile, including the depletion factor in the gap 
(0.24\%), are found to match those of the counter-rotating disk model~1 better than in 
previously considered planet-bearing models. 
Nevertheless, the gap in the planet-bearing disk model still has sharper edges than those
of the counter-rotating disk model.

It appears that for the case of giant planet-bearing, non-self-gravitating disks the gaps tend 
to be deeper than those of the counter-rotating disks unless the viscosity is high 
($\alpha=10^{-2}$) or the planetary mass is low ($\sim1\,M_\mathrm{Jup}$). The density peak at 
the gap outer edge in the planet-bearing model is always higher, implying that a pressure maximum 
is much better expressed in planet-bearing disks than in counter-rotating disks.
In addition, the slope of the planet-bearing disk is usually shallower, than that of the
counter-rotating disk. For instance, the planet-bearing disks in Figure~\ref{fig10} and in
the top panel of Figure~\ref{fig11} have slopes that lies in 
the $\alpha_{\rm d}=-[0.9:1.15]$ limits. The planet-bearing disks in the bottom panel of Figure~\ref{fig11}
and in Figure~\ref{fig12} have somewhat steeper slopes, $\alpha_{\rm d}\approx -1.5$, but still
shallower than those of the counter-rotating disks, $\alpha_{\rm d}=-[1.6:1.8]$. Taking a steeper
initial gas surface density distribution in the planet-bearing disk could 
help to bring the resulting slope in a better agreement with the counter-rotating disk,
but we note that we have already taken a rather steep initial surface density profile,
$\Sigma~r^{-1.5}$, and increasing it even further may be difficult to justify.

In order to check the effect of disk self-gravity, additional simulations were done using 10 
times higher disk mass (similar to that of the counter-rotating model~1) than previously 
considered. We note that in simulations with disk self-gravity the density scaling was 
not applied. For the viscous $\alpha$ 
and planetary mass, we chose $10^{-3}$ and $1.0~M_\mathrm{Jup}$, respectively. As one can see in 
Figure\,\ref{fig13}, the density profile of the inner and outer disks match those of the 
counter-rotating disk model rather well. Significant differences are found only for the innermost 
disk region, where the surface density of the planet-bearing disk has a notable excess, 
and also at the center of the gap, where the planet-bearing disk has weaker depletion 
($2.5\%$). We note that the difference in 
the innermost disk is likely related to different boundary conditions applied in the two numerical codes.

We note that the gap depth and the structure of its edges are expected to be altered if the gas 
thermodynamic is taken into account. Moreover, as the presented snapshots were taken after $1000$
planetary orbits, the orbital migration should also be taken into account presumably affecting 
the shape and the radial position of the gap.

\section{Conclusions}
\label{summary}
In this work we studied numerically the formation and physical properties of
circumstellar disks formed as a result of gravitational collapse of dense 
cloud cores submerged into a low-density external environment counter-rotating 
with respect to the core. We found that counter-rotating disks form during the evolution 
for a wide parameter space of masses and rotation rates in the external environment,  
with the inner disk corotating with the star and the outer disk counter-rotating with 
respect to both the inner disk and the star. 
The inner and outer disks are separated by a deep gap in the gas surface density.
The gap shape, its depth and eccentricity may vary depending on the model. 
The gap often migrates inward but the migration speeds are vastly different, 
ranging from more than a hundred AU per Myr to less than
one AU per Myr, suggesting that this structure may be a 
long-lived phenomenon which is comparable to the lifetime of the disk itself. 

We compared the properties of the gap in counter-rotating disks with those formed as a result of 
gap opening in a giant-planet-bearing disk. 
We found that the shape of the inner disk in both gap-opening mechanisms can be rather similar.
Moreover, given a proper choice
of the planetary mass and viscous $\alpha$-parameter (1.0~$M_{\rm Jup}$ and 10$^{-3}$, respectively) and 
considering relatively-massive (self-gravitating) disks, the shape and the depth of the gap
in planet-bearing models can match those of counter-rotating disks rather well. 
Therefore, the shape of the gas surface density profile in massive planet-bearing disks
may resemble that of counter-rotating disks, which will make it difficult to distinguish
between the two gap-forming mechanisms based solely on the gas density distribution.
This implies that planet-bearing gaps may be confused with gaps formed in counter-rotating disks
and gas kinematic studies are necessary to distinguish between the two possible gap-forming mechanisms.
Another potentially observable effect that can help to distinguish between the two gaps
is the emission due to accretion onto the protoplanet. The H$\alpha$ emission that was recently reported
for the LkCa planetary system \citep{Sallum2015} presents one possible example. The 
differences in the shape of the spiral pattern and the pitch angle may also be used to
determine the origin of the gap, but it requires further investigation and synthetic image 
modeling as was done in, e.g., \citet{Juhasz2015} and \citet{Dong2016}. 

At the same time, gaps in counter-rotating disks are as a rule distinct from those formed in 
planet-bearing models with low-mass (non-self-gravitating) disks.
The latter often possess a deeper gap with sharper edges, implying stronger pressure bumps in the 
vicinity of the gap, which are subject to dust accumulation (see e.g., \citealp{Pinilla2012}). 
Nevertheless, we expect that the peculiar density profile in the outer counter-rotating 
disk can also facilitate the dust growth in these regions, which can lead to an interesting 
perspective of forming a large population of planetesimals and solid protoplanetary cores
counter-rotating to the star and the inner disk.

{\it Acknowledgments.}  We are thankful to the anonymous referee for very useful comments and suggestions
that helped to improve the manuscript.  This project was supported by the Austrian FWF through the NFN project grant
S116 Pathways to Habitability and joint OeAD-OMA program through project 90\"ou25. E.I.V acknowledges support 
from the Russian Ministry of Education and Science Grant 3.961.2014/K. Zs. R. acknowledges support 
from Momentum grant of the MTA CSFK Lend\"ulet Disk Research Group and the Hungarian Grant K101393.
The simulations were performed on the Vienna Scientific Cluster (VSC-2), on the Shared Hierarchical 
Academic Research Computing Network (SHARCNET), on the Atlantic Computational Excellence Network 
(ACEnet). This publication is supported by the Austrian Science Fund (FWF). 
Zs. R. gratefully acknowledges the support of NVIDIA Corporation with the
donation of the Tesla 2075 and K40 GPUs used for this research.

\end{document}